\documentclass[twocolumn,showpacs,preprintnumbers,amsmath,amssymb,prb]{revtex4-1}

\usepackage{graphicx}
\usepackage{dcolumn}
\usepackage{bm}

\begin{document}

\title{Atomistic simulations of structural and thermodynamic properties of bilayer graphene}

\author{K.V. Zakharchenko, J.H. Los, M.I. Katsnelson, A. Fasolino}

\address {
Radboud University Nijmegen, Institute for Molecules and Materials, Heyendaalseweg 135, 6525 AJ Nijmegen, The Netherlands
         }

\date{\today}

\begin{abstract}
We study  the structural and thermodynamic properties of bilayer graphene, a prototype two-layer membrane, by means of Monte Carlo simulations based on the empirical bond order potential LCBOPII.  We present the temperature dependence of lattice parameter, bending rigidity and high temperature heat capacity as well as the correlation function of out-of-plane atomic displacements. The thermal expansion coefficient changes sign from negative to positive above $\approx 400$ K, which is lower than previously found for single layer graphene and close to the experimental value of bulk graphite. The bending rigidity is twice as large than for single layer graphene, making the out-of-plane fluctuations smaller. The crossover from correlated to uncorrelated out-of-plane fluctuations of the two carbon planes occurs for wavevectors shorter than $\approx 3$~nm$^{-1}$.  
\end{abstract}

\maketitle

\section{Introduction}

Bilayer ({\it BL}) graphene has unique electronic properties and its chiral quasiparticles with parabolic dispersion make it different from both single layer ({\it SL}) graphene and bulk graphite~\cite{Novoselov2006}. The energy gap of {\it BL} graphene can be opened and tuned by applying a voltage, with promises for applications~\cite{GeimPRL, Morpurgo}. Also the possibility of some exotic many body phenomena, such as pseudospin magnetism~\cite{McDonald} have been discussed. For these reasons, {\it BL} graphene is currently subject of great interest. However the knowledge of its structural properties is still very poor. It was shown experimentally that, {\it BL} graphene is also corrugated~\cite{SolStComm2007} like {\it SL} graphene, but no systematic study has been carried out. This corrugation (ripples) may constitute an important scattering mechanism for electrons~\cite{PhyslTrans} and ripples can give rise to charge inhomogeneities (electron and hole puddles)~\cite{Polini}. Although important for their relation to electronic properties, the structural properties of  {\it BL} graphene are also important from the point of view of statistical mechanics since the {\it BL} graphene is a unique realization of crystalline membranes formed by two atomic layers. 

Assessing the structure of a {\it BL} graphene is experimentally challenging and theoretical calculations can be particularly helpful. Since the observed corrugations are on a scale much larger than interatomic distances, ab-initio simulations are not feasible. This interesting range of lengths (e.g. for electron interactions with ripples) is, however, not necessarily well described by continuum medium theories~\cite{nelsonbook}. Atomistic simulations based on accurate empirical interaction potentials are particularly suitable for this purpose. We have recently studied the structural and thermodynamic properties of {\it SL} graphene~\cite{NatMat, kostya1, los2} by Monte Carlo ({\it MC}) simulations  based on the LCBOPII bond order potential~\cite{los1}. Here we present the results of similar calculations for {\it BL} graphene, where a new aspect related to the correlation of atomic displacements in different layers arises.

\section{Method of calculation}

We perform {\it MC} simulations in the $NPT$ ensemble at pressure $P=0$ and temperature $T$  with periodic boundary conditions for samples of $N=16128$ and $N=8640$ atoms per layer. When not specified, the presented results are for the largest sample. The equilibrium size at $T=0$~K of the $N=16128$ sample is $L_x=20.66$~nm in $x$ and $L_y=20.448$~nm in $y$ direction and that of the $N=8640$ sample is $L_x=14.757$~nm and $L_y=15.336$~nm. The finite size of our sample defines the lowest accessible wavevectors is $x$ and $y$ directions as $q_x=2\pi/L_x$ and $q_y=2\pi/L_y$.
Motivated by the results of recent quantum {\it MC} calculations~\cite{spanu}, we have slightly modified the long-range part of LCBOPII as to have an interlayer binding energy of $50$~meV/atom against the $25$~meV/atom of the parametrization of Ref.~\onlinecite{los1}, while keeping the interlayer compressibility constant. 

We equilibrate the sample for at least $5\cdot10^5$ steps ($1$ {\it MC} step corresponds to $N$ attempts to a coordinate change), using the recently introduced {\it MC} sampling based on collective atomic moves (wave moves)~\cite{los2} in addition to conventional {\it MC} moves. This technique was successfully introduced for {\it SL} graphene. For {\it BL} graphene it was extended as follows. Wave moves are applied to both layers simultaneously, or only to the upper or lower layer, with equal probabilities for the three cases. The amplitude $A_1$ of the wave moves applied to both layers simultaneously is different from the amplitude $A_2$ of the wave moves applied to either upper or lower layer separately. The amplitudes $A_1$ and $A_2$ are chosen in such a way that the acceptance rate for wave moves is between $0.4$ and $0.5$ for any of these three cases.

Further $5\cdot10^5$ {\it MC} steps are used to evaluate the temperature dependence of the ensemble averages.

\section{Results}

\begin{figure}
\includegraphics[clip=true,width=0.9\linewidth]{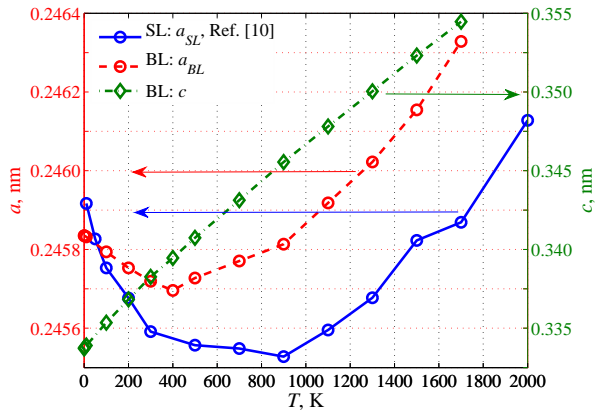}
\caption{(color online) Temperature dependence of the in-plane lattice parameter $a$ of {\it SL} (circles, solid, blue, from Ref.~\onlinecite{kostya1}) and {\it BL} (circles, dashed, red) graphene, and of the interlayer distance $c$ of {\it BL} graphene (diamonds, dash-dotted, green). At $T=0$, $a_{SL}=0.24595$~nm, $a_{BL}=0.24583$~nm, $c=0.33371$~nm.}
\label{fig1}
\end{figure}

The temperature dependence of the in-plane lattice parameter $a$ and of the interlayer distance $c$ of {\it BL} graphene are shown in Fig.~\ref{fig1}. The in-plane lattice parameter $a$ of {\it BL} graphene decreases with increasing temperature up to about 400~K, yielding a negative thermal expansion coefficient $\alpha_a = d\ln{a}/dT = (-3.0\pm0.7)\cdot10^{-6}~{\rm K}^{-1}$ in the range 0--300~K. The behavior of $a(T)$ differs from that of {\it SL} graphene, which has a minimum of $a$ at $T \approx 900~K$ and $\alpha_a = (-4.8\pm1.0)\cdot10^{-6}~{\rm K}^{-1}$~(see Ref.~\onlinecite{kostya1}) in the range 0--300~K, and is similar to bulk graphite, which has a minimum of $a$ between $300$ and $500$~K~\cite{MoMar, NelsonRiley}. We note that our approach is classical and therefore not appropriate in the low temperature limit. However, since the thermal expansion is mostly determined by the low-frequency bending modes~\cite{MoMar}, a classical description is already justified below room temperature. Indeed for single layer graphene, our results for $a(T)$ between $100$~K and $400$~K agree very well with those of Ref.~\onlinecite{MoMar} where the quantum statistics of phonons was taken into account. 

In Ref.~\onlinecite{MoMar}, the temperature dependence of $a$ for {\it SL} graphene and bulk graphite has been determined in the quasiharmonic approximation with phonon frequencies and Gruneisen parameters calculated from first principles. While for the case of bulk graphite these calculations reproduce the non monotonic behavior of $a(T)$ observed experimentally, for {\it SL} graphene $a(T)$ keeps decreasing up to high temperatures. In our simulations of {\it SL} graphene~\cite{kostya1} we found instead a non monotonic behavior of $a(T)$. The experimental value of $a(T)$ for {\it SL} graphene that was measured up to $400$~K~\cite{GrapheneAlphaExp} seems to support our results.

The discrepancy with quasiharmonic results should be due to the fact that this method~\cite{MoMar} neglects self-anharmonic effects~\cite{katsnelson2005Enc}, namely multiphonon contributions to the free energy. Of course, in our simulations, the thermal expansion is calculated directly and {\it all} anharmonic effects are taken into account. Unfortunately, we do not have results for  bulk graphite with the same in-plane area, due to the long range part of our potential that, with a cut off of $0.6$~nm, requires to simulate samples with at least four layers with periodic boundary conditions. Nevertheless, we believe that the fact that the thermal expansion of {\it BL} graphene is similar to the one resulting from  quasiharmonic theory for bulk graphite suggests that multiphonon processes are much less important in {\it BL} graphene, compared to {\it SL} graphene. 

In Fig.~\ref{fig1} we also show the interlayer distance $c$ that grows with temperature, similarly to bulk graphite~\cite{MoMar}, with an out-of-plane thermal expansion coefficient $\alpha_c  = d\ln{c}/dT =(3.5\pm0.5)\cdot10^{-5}~{\rm K}^{-1}$, which is comparable to the experimental value for bulk graphite, $\alpha_c=2.7\cdot10^{-5}~{\rm K}^{-1}$ (see Ref.~\onlinecite{NelsonRiley}).

\begin{figure}
\rotatebox{270}{\includegraphics[clip=true,width=0.7\linewidth]{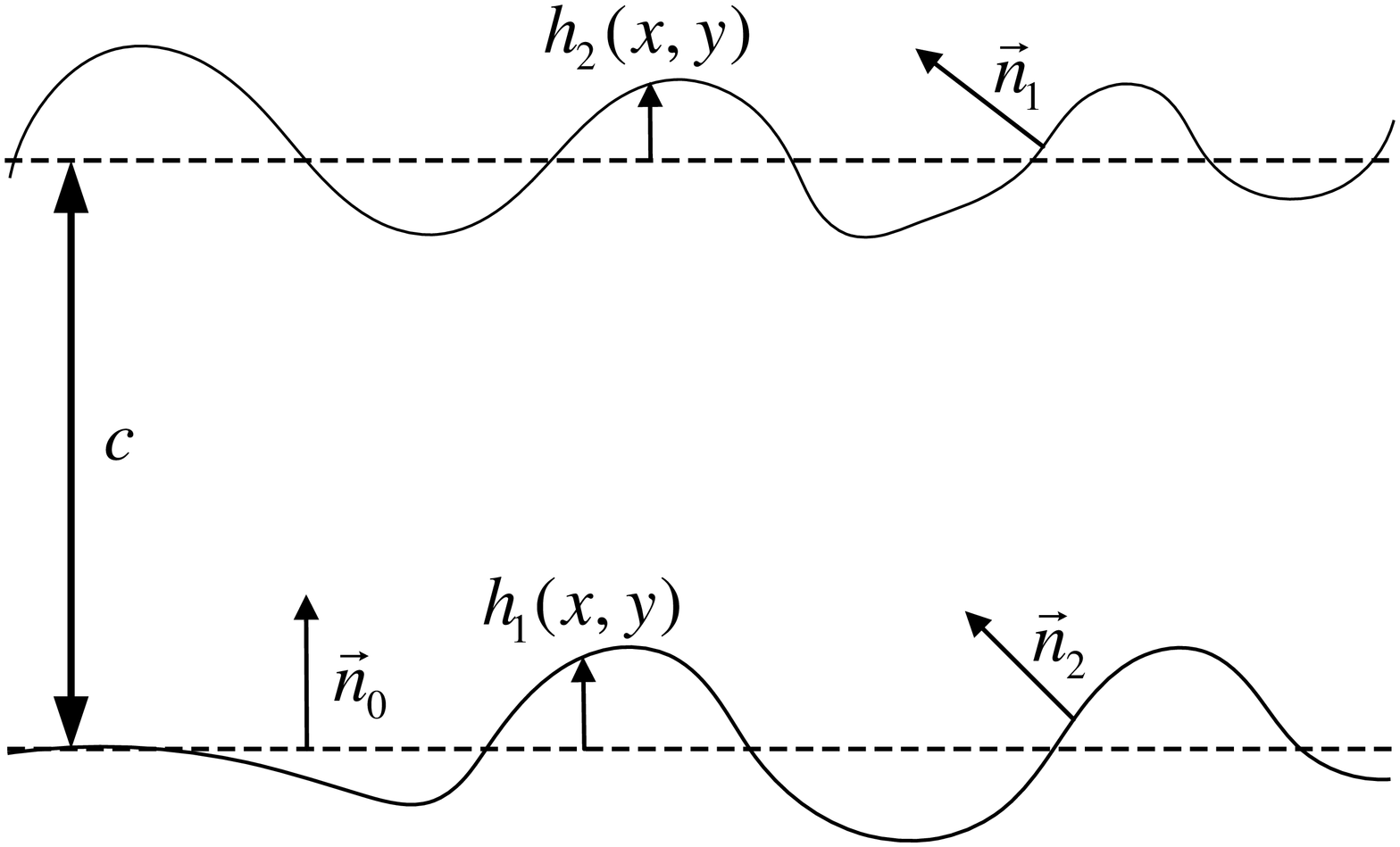}}
\caption{Schematic view of {\it BL} graphene (solid lines). $h_1$ and $h_2$ are out-of-plane  deviations with respect to the middle planes (dashed lines). The unit vectors $\vec{n}_1$ and $\vec{n}_2$ are the normals to each point in the upper and lower layer respectively. $\vec{n}_0$ is the normal to the reference plane. $c$ is the interlayer distance. The figure is schematic and does not show the real scale of the fluctuations.}
\label{fig2}
\end{figure}

We now proceed to a study of thermal bending fluctuations. In the continuum limit graphene can be described as a flexible crystalline membrane~\cite{nelsonbook, NatMat, los2} which is characterized by a two component in-plane phonon field $u_\alpha(\vec x), \alpha=1,2$ and a one component out-of-plane displacement field $h(\vec x)$. The effective free energy is given by the sum of bending energy and in-plane elastic energy~\cite{nelsonbook}
\begin{equation}
\label{eq:Free_F_1}
{\cal H} = \frac{1}{2}\int\! d^2\!x\,\left(\kappa \left(\nabla^2 h\right)^2 + \mu u_{\alpha\beta}^2+\frac{\lambda}{2}u_{\alpha\alpha}^2\right),
\end{equation}
where the strain tensor $u_{\alpha\beta}$ is
\begin{equation}
\label{eq:strain}
u_{\alpha\beta} = \frac{1}{2}\left(\partial_{\alpha}u_{\beta}+\partial_{\beta}u_{\alpha}+\partial_{\alpha}\!{h}\,\partial_{\beta}\!{h}\right),
\end{equation}
$\kappa$ is the bending rigidity and  $\mu$ and $\lambda$ are  Lam\'{e} coefficients.

In first approximation, {\it BL} graphene can be considered as two {\it SL} graphene layers interacting with each other. The natural way to describe {\it BL} graphene, is to use the out-of-plane  deviations from the center of mass of each layer, $h_1$ and $h_2$ for upper and lower layer respectively as sketched in Fig.~\ref{fig2}. Thus, {\it BL} graphene can be parametrised by the average height fluctuation field $h=\left(h_1+h_2\right)/2$ and thickness fluctuation field $\delta h=h_1-h_2$. 
  
The part of the Hamiltonian~(\ref{eq:Free_F_1}) related to out-of-plane displacements can thus be written as:
\begin{equation}
\label{eq:Free_F_4}
{\cal H}_{\rm out} = \frac{1}{2}\int\! d^2\!x\,\left(\kappa \left(\nabla^2 h_1\right)^2 + \kappa\left(\nabla^2 h_2\right)^2+ 2\gamma \left( \delta h \right) ^2 \right),
\end{equation}
where the first two terms are responsible for the bending energy of the upper and lower layers and $\kappa$ is the bending rigidity per layer. We have introduced the last term characterized by the parameter $\gamma$ to account for interlayer interactions. Substituting $h_1$ and $h_2$ with $h\pm\delta h/2$, we obtain:
\begin{equation}
\label{eq:Free_F_5}
{\cal H}_{\rm out} = \frac{1}{2}\int\! d^2\!x\,\left(2\kappa \left(\nabla^2 h\right)^2 + \frac{\kappa}{2}\left(\nabla^2 \delta h\right)^2 + 2\gamma \left(\delta h\right)^2\right).
\end{equation}

In the harmonic approximation, which means neglecting the last term of the strain tensor~(\ref{eq:strain}), the out-of-plane  $h(x)$ and in-plane  $u_\alpha(x)$ modes are decoupled. In this approximation, the mean square Fourier components of the field $h(\vec q)$ with wavevectors $\vec q$ are:
\begin{equation} 
\label{eq:h_q_eq1}
\langle |h(\vec q)|^2 \rangle=\frac{N}{S_0}\frac{T}{2\kappa q^4},
\end{equation}
and of the field  $\delta h (\vec q)$ are:
\begin{equation} 
\label{eq:h_q_eq2}
\langle |\delta h(\vec q)|^2 \rangle=\frac{N}{S_0}\frac{T}{\frac{1}{2}\kappa q^4 + 2\gamma},
\end{equation}
where $N$ is the number of atoms per layer and $S_0$ is the area per atom in the layer. If the bending rigidity of a {\it SL} graphene is the same as the bending rigidity per layer of {\it BL} graphene, then it follows from Eq.~(\ref{eq:h_q_eq1}), that  $\langle |h(\vec q)|^2 \rangle$ for {\it BL} graphene is twice smaller than for {\it SL} graphene. This is actually a very good approximation as we will show below. 

We further introduce the notation $H(q) \equiv \langle |h(\vec q)|^2 \rangle$ and $\Delta H(q) \equiv \langle |\delta h(\vec q)|^2 \rangle$.

An alternative way to describe out-of-plane fluctuations is via the unit vector normal to the average surface between two layers:
\begin{equation}
\label{eq:G_q_eq1}
n_i(\vec{x})=-\frac{\partial_i{h}} {\sqrt{1+|\nabla h|^2}},
\end{equation}
with $i=1,2$~\cite{nelsonbook}. 

The correlation function of the normals, $G(q)= \langle |\vec n(\vec q)|^2 \rangle$ is equal to $q^2H(q)$ if $|\nabla h|^2 \ll 1$.  Thus, in the harmonic approximation
\begin{equation} 
 \label{eq:g_q_eq2}
G(q) = \frac{N}{S_0}\frac{T}{2 \kappa q^2}.
\end{equation}
which is a factor 2 smaller than $G(q)$ in {\it SL} graphene~\cite{NatMat, los2}.

The correlation functions $H(q)$ and $G(q)$ are calculated independently as described below.  In principle to calculate $H(q)$, we have to calculate the Fourier transforms of the atomic displacements  $h(\vec{x})$. However, the atomic positions in a generic configuration in {\it MC} simulations are discontinuous and  should be smoothed. This problem is related to the numerical calculations of derivatives and different operators on the hexagonal lattice~\cite{zakrzewsky}. Our procedure is the following. Let $h_0$ be the $z$-coordinate of an atom and $h_a$, $h_b$ and $h_c$ the $z$-coordinates of its three nearest neighbors. Then  the averaged out-of-plane displacement of the central atom $\widetilde{h_0}$ is defined as:
\begin{equation} 
 \label{eq:h_smooth}
\widetilde{h_0} = \frac{1}{2}\left(h_0+\frac{1}{3}\left(h_a+h_b+h_c\right)\right).
\end{equation}

This value is used to calculate the Fourier components $h(\vec q)$  using the wave vectors defined by periodic boundary conditions of the undistorted lattice. The normals needed to calculate $G(q)$, instead, are automatically smooth because they are calculated as averages of the normals to the three planes defined by three vectors, connecting the central atom to its three nearest neighbors~\cite{NatMat, los2}. For {\it BL} graphene, we calculate the correlation function $G(q)$ for the normals of all atoms in the two layers. 

\begin{figure}
\includegraphics[width=0.9\linewidth]{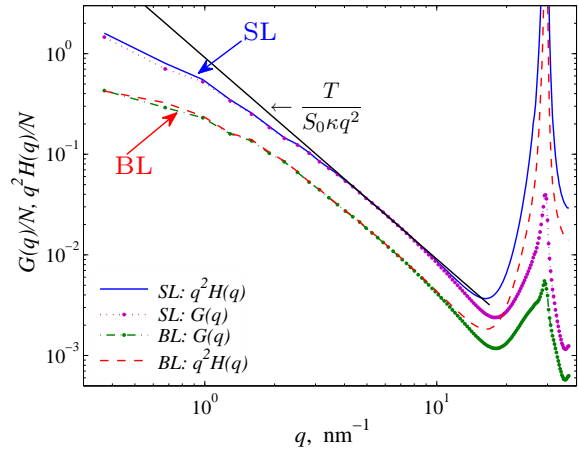}
\caption{(color online) Normal-normal correlation function  $G(q)/N$ at $T=300$ K for {\it SL} (solid blue line) and {\it BL} (dashed red line) graphene compared to $q^2H(q)$ for single (dotted magenta line) and {\it BL} (dash-dotted red line) graphene. The solid black straight line shows the fit ($\sim q^{-2}$) in the harmonic part.}
\label{fig3}
\end{figure}

Fig.~\ref{fig3} shows the correlation functions $G(q)/N$ and $q^2 H(q)/N$ for {\it SL} and {\it BL} graphene at $T=300$~K. We plot these functions as a function of $q=|\vec q|$ by giving their average value at all allowed wavevectors with the same modulus. The difference between $G(q)/N$ and $q^2 H(q)/N$ is negligible for $q<10$~nm$^{-1}$ where the condition $|\nabla h|^2 \ll 1$ is satisfied. The functions $H(q)$ and $G(q)$ behave according to the harmonic approximation Eqs.~(\ref{eq:h_q_eq1}) and~(\ref{eq:g_q_eq2}) for $q$ from $3$ to $9$ nm$^{-1}$ as it  is also shown in Fig.~\ref{fig3}. In this interval the correlation functions for {\it BL} graphene are about twice smaller than for {\it SL} graphene, which means that the effective bending rigidity for {\it BL} graphene is twice larger than the one of {\it SL} graphene, as we had guessed above. The deviation from the harmonic approximation for $q<3$~nm$^{-1}$ is due to the coupling between bending and stretching modes in Eq.~(\ref{eq:strain})~\cite{nelsonbook}.

\begin{figure}
\includegraphics[width=0.9\linewidth]{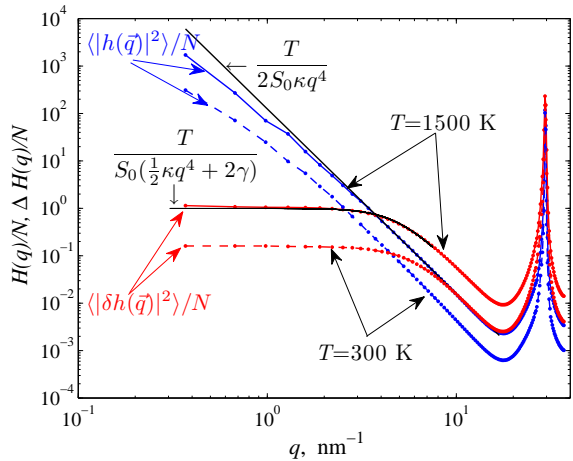}
\caption{(color online) Average height $\langle |h(\vec q)|^2 \rangle$ (blue lines) and thickness $\langle |\delta h(\vec q)|^2 \rangle$ (red lines) fluctuations of {\it BL} graphene at $T=300$~K (dashed lines) and $T=1500$~K (solid lines). Black solid lines show the fit according to the Eqs.~(\ref{eq:h_q_eq1})--(\ref{eq:h_q_eq2}).}
\label{fig4}
\end{figure}

Fig.~\ref{fig4} shows the correlation functions $H(q)/N$ and $\Delta H(q)/N$ of {\it BL} graphene for $T=300$~K and $T=1500$~K together with the harmonic fit according to Eqs.~(\ref{eq:h_q_eq1})--(\ref{eq:h_q_eq2}). The function $\Delta H(q)$ is specific of a bilayer and has no analog for single layer membranes. First of all, we note that $\Delta H(q)$ follows the harmonic approximation Eq.~(\ref{eq:h_q_eq2}) in the whole studied range of $q$, even where the deviations of $H(q)$ from the harmonic approximation Eq.~(\ref{eq:h_q_eq1}) are pronounced. This means that thickness fluctuations are much less coupled to in-plane fluctuations, than average out-of-plane fluctuations.

The second noticeable point, is the $q$-independent behavior of $\Delta H(q)$ for $q<q^*\approx 3$ nm$^{-1}$. In turn, this means that, in this range of $q$, the out-of-plane fluctuations of the two carbon layers are strongly coupled and only one soft mode $h(q)$ survives.  Therefore, at scales larger than $2\pi/q^*\approx 2$~nm, {\it BL} graphene can be considered as a single membrane, whereas at smaller scales, $h_1(q)$ and $h_2(q)$ fluctuate rather independently. Indeed it follows from Eqs.~(\ref{eq:h_q_eq1}), (\ref{eq:h_q_eq2}) that if one neglects the interlayer coupling $\gamma$ in Eq.~(\ref{eq:h_q_eq2}) one has: 
\begin{equation}
\label{eq:appr}
\langle h_1(q)h_2(q)\rangle = 0 , \langle|h_1(q)|^2\rangle=\langle|h_2(q)|^2\rangle=\frac{N}{S_0} \frac{T}{\kappa q^4}.
\end{equation}

In general, the perfect coincidence of $\Delta H(q)$ calculated from the {\it MC} simulations, with the theoretical prediction~(\ref{eq:h_q_eq2}) of the Hamiltonian~(\ref{eq:Free_F_5}) confirms the correct choice of Hamiltonian to describe {\it BL} graphene.

The crossover at $q^*$ from independent to coherent fluctuations in the two layers is important for the scattering of electrons from height fluctuations in {\it BL} graphene, which is determined mainly by long range fluctuations with strongly $q$-dependent correlation functions (compare with Ref.~\onlinecite{PhyslTrans} for {\it SL} graphene). Therefore, fluctuations of the interlayer distance become irrelevant for electrons with wavevector $k<q^*$.  Moreover, for sample sizes $L>2\pi/q^* \approx 2$~nm the height fluctuations in {\it BL} graphene are expected to be weaker than in {\it SL} graphene, because in the regime of coherent fluctuations the bilayer is twice stiffer than a single layer. This is qualitatively confirmed by the results presented in Fig.~\ref{fig6} where we compare the values of $\langle h^2 \rangle$ for {\it SL} and {\it BL} graphene. 

\begin{figure}
\includegraphics[width=0.9\linewidth]{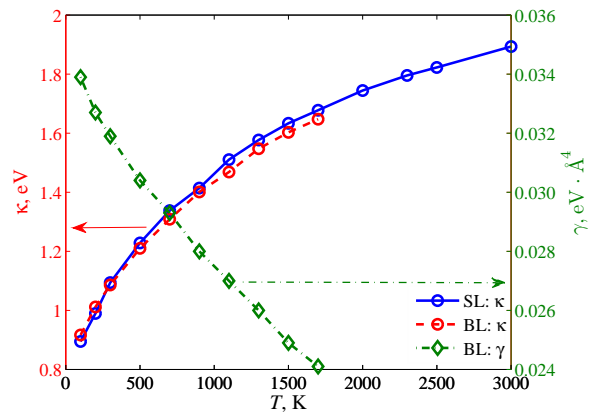}
\caption{(color online) Temperature dependence of the bending rigidity $\kappa$ of {\it SL} graphene (circles, solid, blue), bending rigidity per layer $\kappa$ of {\it BL} graphene (circles, dashed, red)  and parameter $\gamma$ of {\it BL} graphene (diamonds, dash-dotted, green).}
\label{fig5}
\end{figure}

The temperature dependence of the parameters $\kappa$ and $\gamma$ of {\it BL} graphene are presented in Fig.~\ref{fig5} together with the parameter $\kappa$ of {\it SL} graphene. The parameter $\gamma$ of {\it BL} graphene decreases with temperature, which is not surprising. This parameter is responsible for the interlayer coupling, and it decreases with temperature since the interlayer distance $c$ increases with temperature (Fig.~\ref{fig1}). The effective bending rigidity $\kappa$ grows with temperature in agreement with the general theory of crystalline membranes~\cite{NelsonPeliti}, as well as with our previous numerical results for {\it SL} graphene~\cite{NatMat, los2}. The behavior of liquid membranes is known to be opposite, with $\kappa$ decreasing with temperature~\cite{Leibler}. The statement that $\kappa$ decreases with $T$ also for graphene~\cite{singapur} is therefore in disagreement with general arguments~\cite{nelsonbook, NelsonPeliti} and our results. The point is that the origin of the main anharmonic effects in liquid and crystalline membranes are completely different. For liquid membranes they originate from high order terms of the mean curvature in $\nabla h$, which results in perturbative corrections to $\kappa$ that are of the form $T\ln{qa}<0$ with $a$ the interatomic distance~\cite{NelsonPeliti, Leibler}. For crystalline membranes, instead, perturbative corrections to $\kappa(T)$ due to the coupling of bending and out-of-plane fluctuations are much stronger, positive and proportional to $T/q^2$ (Ref.~\onlinecite{NelsonPeliti}). 

Actually the fact that $d\kappa/dT>0$ for crystalline membranes has a very simple meaning: as the temperature increases, the amplitude of corrugation also increases, resulting in a strengthening of the membrane~\cite{Landau}. As already mentioned, the bending rigidity per layer of {\it BL} graphene turns out to be very close to that of {\it SL} graphene, which is not surprising since the interlayer coupling is much weaker than the in-plane chemical bonding. However, since the renormalization of $\kappa$ is strongly $q$-dependent for crystalline membranes, the definition of $\kappa(T)$ should be further specified. What is shown as $\kappa(T)$ in Fig.~\ref{fig5}, and what was previously calculated for {\it SL} graphene in Refs.~\cite{NatMat, los2} are the results of a best fit of the correlation functions $G(q)$ and $H(q)$ in the $q$ range where the slope can be well approximated by the harmonic behavior of Eqs.~(\ref{eq:Free_F_5}) and~(\ref{eq:g_q_eq2}). Since, in this interval of $q$, the out-of-plane fluctuations of either layer of {\it BL} graphene are of the same order as those of {\it SL} graphene (see Eq.~(\ref{eq:appr})), it is not surprising that the temperature dependence of $\kappa$ for {\it BL} graphene is only marginally smaller than for the one of {\it SL} graphene.

It is important to notice, however, that the macroscopic behavior of the bending rigidity of free membranes for $q\rightarrow 0$ at finite temperature is divergent as $q^{-\eta}$ with $\eta \approx 0.85$ (see Refs.~\onlinecite{nelsonbook},~\onlinecite{los2}). The size of the {\it BL} graphene samples used here makes an estimate of $\eta$ for this case not precise enough as to be compared quantitatively to that found for {\it SL} graphene~\cite{los2}, but the qualitative behavior shown in Fig.~\ref{fig3} is very similar for {\it SL} and {\it BL} graphene.

\begin{figure}
\includegraphics[width=0.9\linewidth]{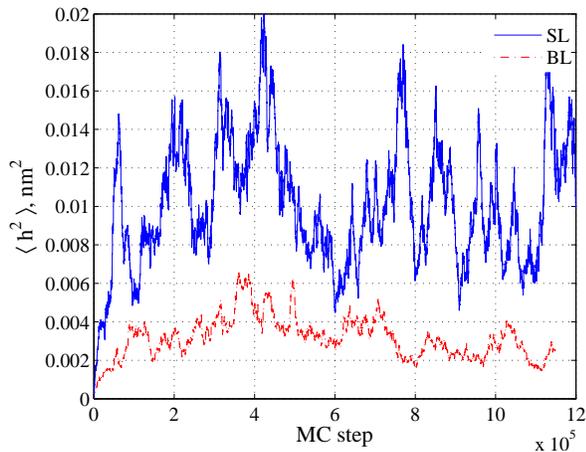}
\caption{(color online) Height fluctuations of {\it SL} graphene (solid blue line) compared to one of {\it BL} graphene (dashed red line) as a function of {\it MC} step at $T=300$ K.}
\label{fig6}
\end{figure}

The mean square height fluctuations $\langle h^2 \rangle=\sum_{q} H(q)$ are equally size-dependent. Since the sum over $q$ is divergent at the lower limit $q_{min}=2\pi/L$, $\langle h^2 \rangle$  is mostly determined by the effective $\kappa(q)$ for the smallest wavevectors and therefore, for large enough samples, it should scale as $L^{2-\eta}$ (see Refs.~\onlinecite{nelsonbook},~\onlinecite{los2}). According to Fig.~\ref{fig4}, deviations of 
$H(q)$ from harmonic behavior occur for $q < 1$ nm$^{-1}$ and thus, the crossover from harmonic behavior $h^2 \propto L^2$ to the anharmonic one 
$h^2 \propto L^{2-\eta}$ takes place for sample size $L \approx 6$~nm.

\begin{figure}
\includegraphics[width=0.9\linewidth]{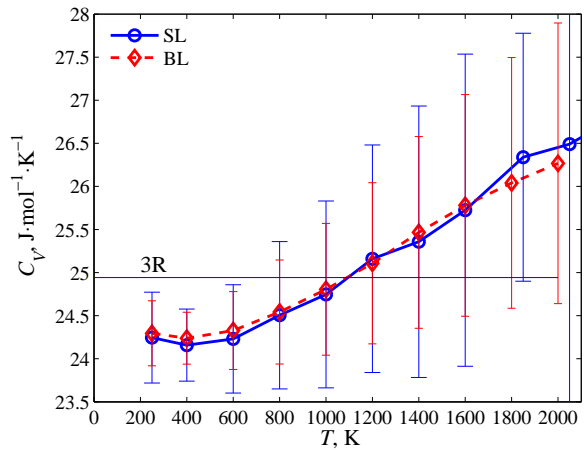}
\caption{(color online) Temperature dependence of the molar heat capacity at constant volume $C_V$ of {\it SL} (solid blue line) and {\it BL} (dashed red line) graphene. Data obtained for $N=8640$ atoms sample.}
\label{fig7}
\end{figure}

To characterize qualitatively the anharmonicity at the atomic scale, we calculate the temperature dependence of the molar heat capacity at constant volume
\begin{equation}
C_V=\frac{3R}{2}+\frac{dU}{dT},
\end{equation}
where $U$ is the potential energy and $R$ the gas constant. In Fig.~\ref{fig7} we compare the results with those of {\it SL} graphene~\cite{kostya1}.

Three and four phonon processes result in the linear growth of $C_V$ at high temperatures~\cite{katsnelson2005Enc}. One can see that {\it SL} and {\it BL} graphene are almost the same as expected, since phonons of the whole Brillouin zone contribute to this quantity and the phonon spectra of {\it SL} and {\it BL} graphene differ only slightly close to the $\Gamma$ point (see, e.g., the calculated phonon spectra of graphene and graphite in Ref.~\onlinecite{MoMar}). 

\section{Summary}

In conclusions, we have studied several temperature dependent properties of {\it BL} graphene by means of classical {\it MC} simulations. The high temperature heat capacity is similar to that of {\it SL} graphene, whereas the thermal expansion is essentially different and close to the one experimentally observed in graphite.

We also introduced a new Hamiltonian which accounts for interlayer interactions in {\it BL}  graphene and showed that it correctly describes the behavior of {\it BL} graphene. We have found that, depending on the wavevector, the height fluctuations in the two layers are either coherent(for $q<q^*$) or incoherent (for $q>q^*$) with $q^* \approx 3$~nm$^{-1}$ at room temperature and we have discussed the consequences of this fact for observable properties, like height fluctuations and  electron scattering.

\acknowledgments
This work is part of the research program of the Stichting voor Fundamenteel Onderzoek der Materie (FOM), which is financially supported by the Nederlandse Organisatie voor Wetenschappelijk Onderzoek (NWO). We thank Marco Polini and Nils Hasselmann for discussions.


\begin{thebibliography}{99}
\bibitem{Novoselov2006} K. S. Novoselov, E. McCann, S. V. Morozov, V. I. Fal’ko, M. I. Katsnelson, U. Zeitler, D. Jiang, F. Schedin, A. K. Geim, Nature Phys. {\bf 2}, 177 (2006).
\bibitem{GeimPRL} E. V. Castro, K. S. Novoselov, S. V. Morozov, N. M. R. Peres, J. M. B. Lopes dos Santos, J. Nilsson, F. Guinea, A. K. Geim, and A. H. Castro Neto, Phys. Rev. Lett. {\bf 99}, 216802 (2007).
\bibitem{Morpurgo}  J. B. Oostinga, H. B. Heersche, X. Liu, A. F. Morpurgo, L. M. K. Vandersypen, Nature Mater. {\bf 7}, 151 (2007).
\bibitem{McDonald} H. Min, G. Borghi, M. Polini, and A. H. MacDonald, Phys. Rev. B {\bf 77}, 041407 (R) (2008). 
\bibitem{SolStComm2007} J. C. Meyer, A. K. Geim, M. I. Katsnelson, K. S. Novoselov, D. Obergfell, S. Roth, C. Girit, and A. Zettl, Solid State Commun. {\bf 143}, 101 (2007).
\bibitem{PhyslTrans} M. I. Katsnelson and A. K. Geim, Phil. Trans. R. Soc. A {\bf 366}, 195 (2008).
\bibitem{Polini} M. Gibertini, A. Tomadin, M. Polini, A. Fasolino, and M. I. Katsnelson, Phys. Rev. B, accepted for publication.
\bibitem{nelsonbook} D. R. Nelson,  T. Piran, and S. Weinberg (Eds), \textit{Statistical Mechanics of Membranes an Surfaces} (World Scientific, Singapore, 2004), ch. 6 and ch. 11.
\bibitem{NatMat} A. Fasolino, J. H. Los and M. I. Katsnelson, Nature Mater. {\bf 6}, 858 (2007).
\bibitem{kostya1} K. V. Zakharchenko, M. I. Katsnelson, and A. Fasolino, Phys. Rev. Lett.  {\bf 102}, 046808 (2009).
\bibitem{los2} J. H. Los, M. I. Katsnelson, O. V. Yazyev, K. V. Zakharchenko and A. Fasolino, Phys. Rev. B  {\bf 80}, 121405 (R) (2009).

\bibitem{los1} J. H. Los, L. M. Ghiringhelli, E. J. Meijer and A. Fasolino, Phys. Rev. B  {\bf 72}, 214102 (2005).
\bibitem{spanu} 
L. Spanu, S. Sorella, and G. Galli,
Phys. Rev. Lett. {\bf 103}, 196401 (2009) and references therein.
\bibitem{MoMar} N. Mounet and N. Marzari, Phys. Rev. B \textbf{71}, 205214 (2005).
\bibitem{katsnelson2005Enc} M. I. Katsnelson, {\it Encyclopedia of Condensed Matter Physics}, ed. by G. F. Bassani, G. L. Liedl, and P. Wyder (Elsevier, Amsterdam, 2005), p. 77.
\bibitem{NelsonRiley} J. B. Nelson and D. P. Riley, Proc. Phys. Soc. \textbf{57}, 477 (1945).
\bibitem{GrapheneAlphaExp} W. Bao, F. Miao, Z. Chen, H. Zhang, W. Jang, C. Dames and C. N. Lau, Nature Nanotech. {\bf 4}, 562 (2009).
\bibitem{zakrzewsky} W. J. Zakrzewski, J. Nonlinear Math. Phys. {\bf 12}, 530 (2005).
\bibitem{NelsonPeliti} D. R. Nelson and L. Peliti, J. Physique \textbf{48}, 1085 (1987).
\bibitem{Leibler} L. Peliti and S. Leibler, Phys. Rev. Lett. {\bf 54}, 1690  (1985). 
\bibitem{singapur} P. Liu and Y. W. Zhang, Appl. Phys. Lett. {\bf 94}, 231912 (2009).
\bibitem{Landau} L. D. Landau and E. M. Lifshitz, \textit{Theory of Elasticity} (Pergamon Press, New York, 1959).
\end{thebibliography}
\end{document}